\documentclass[%
 reprint,
 amsmath,amssymb,
prmaterials,
]{revtex4-2}

\usepackage{graphicx}
\usepackage{dcolumn}
\usepackage{bm}
\graphicspath{{Images/}}
\DeclareUnicodeCharacter{2212}{-}

\begin{document}

\preprint{APS/123-QED}

\title{Understanding Cu Incorporation in the $\mathrm{Cu_{2x}Hg_{2-x}GeTe_4}$ Structure using Resonant X-ray Diffraction}

\author{Ben L. Levy-Wendt}
 \altaffiliation[Also at ]{Stanford University, Stanford, California, 94305 USA.}
\author{Donata Passarello}
\author{Kevin H. Stone}
\author{Michael F. Toney}
 \email{mftoney@slac.stanford.edu}
 \altaffiliation[Also at ]{University of Colorado Boulder, Boulder, Colorado 80309, USA}
\affiliation{%
 SLAC National Accelerator Laboratory, Menlo Park, California, 94025 USA
}%

\author{L\'{i}dia C. Gomes}
 \altaffiliation[Also at ]{Instituto de F\'isica Te\'orica, S\~ao Paulo State University (UNESP), S\~ao Paulo, Brazil}
\author{Elif Ertekin}
 \email{ertekin@illinois.edu}
\affiliation{
 University of Illinois at Urbana-Champaign, Urbana, Illinois 61820, USA
}%
\affiliation{
Naional Center for Supercomputing Applications, Urbana, Illinois 61801, USA
}%

\author{Brenden R. Ortiz}
 \altaffiliation[Also at ]{University of California Santa Barbara, Santa Barbara, California 93106, USA}
\author{Eric S. Toberer}
 \email{etoberer@mines.edu}
\affiliation{%
Colorado School of Mines, Golden, Colorado 80401, USA
}%

\collaboration{DMREF Collaboration}

\date{\today}

\begin{abstract}
The ability to control carrier concentration based on the extent of Cu solubility in the $\mathrm{Cu_{2x}Hg_{2-x}GeTe_4}$ alloy compound (where 0 $\leq$ x $\leq$ 1) makes $\mathrm{Cu_{2x}Hg_{2-x}GeTe_4}$ an interesting case study in the field of thermoelectrics. While Cu clearly plays a role in this process, it is unknown exactly how Cu incorporates into the $\mathrm{Cu_{2x}Hg_{2-x}GeTe_4}$ crystal structure and how this affects the carrier concentration. In this work, we use a combination of resonant energy X-ray diffraction (REXD) experiments and density functional theory (DFT) calculations to elucidate the nature of Cu incorporation into the $\mathrm{Cu_{2x}Hg_{2-x}GeTe_4}$ structure. REXD across the $\mathrm{Cu_k}$ edge facilitates the characterization of Cu incorporation in the $\mathrm{Cu_{2x}Hg_{2-x}GeTe_4}$ alloy and enables direct quantification of anti-site defects. We find that Cu substitutes for Hg at a 2:1 ratio, wherein Cu annihilates a vacancy and swaps with a Hg atom. DFT calculations confirm this result and further reveal that the incorporation of Cu occurs preferentially on one of the z = 1/4 or z = 3/4 planes before filling the other plane. Furthermore, the amount of $\mathrm{Cu_{Hg}}$ anti-site defects quantified by REXD was found to be directly proportional to the experimentally measured hole concentration, indicating that the $\mathrm{Cu_{Hg}}$ defects are the driving force for tuning carrier concentration in the $\mathrm{Cu_{2x}Hg_{2-x}GeTe_4}$ alloy. The link uncovered here between crystal structure, or more specifically anti-site defects, and carrier concentration can be extended to similar cation-disordered material systems and will aid the development of improved thermoelectric and other functional materials through defect engineering. 
\end{abstract}

\maketitle


\section{\label{sec:level1}Introduction}
Cation-disordered semiconductors are an emerging class of materials with properties that are largely governed by lattice site disorder. Site disorder and occupancy are particularly important in optoelectronic and thermoelectric materials, where defect engineering can be utilized to design materials with improved properties such as carrier concentration or lattice thermal conductivity \cite{doi:10.1021/acsenergylett.0c00576, Pan2020}. Further insight on how cation-disorder can be utilized to control a material's properties depends on establishing detailed structure-function relationships. However, pinpointing the source of disorder within materials as well as quantifying the extent of disorder are experimentally challenging tasks.  

Quantitative insight into structure-property relationships in cation-disordered materials is often missing from an experimental viewpoint due to the difficulty of characterizing  small defect/dopant concentrations. Traditional methods of structural characterization, such as X-ray diffraction (XRD), are excellent for understanding a material's long range order but are notoriously poor at quantifying small concentrations of defects and dopants. To address this deficiency, a variant of XRD --- known as resonant energy X-ray diffraction (REXD) --- takes advantage of the energy dependence of atomic scattering factors by measuring the intensity of Bragg peaks as a function of energy across an elemental absorption edge. The atomic scattering factor of a given element changes significantly near its respective absorption edge, which enables the ability to tune the scattering effects of a given element. The change in scattering power elucidates lattice site occupancies that are difficult to identify with traditional XRD, such as elements with similar atomic number occupying symmetric sites \cite{Rekha, doi:10.1021/ja063695y}. REXD can also be used to probe the presence of elements in small point defect quantities \cite{Kevin, Laura, Dmitrienko:av0029}.  Here, we use REXD to examine cation site disorder in thermoelectric materials, which require a delicate optimization between low thermal conductivity and high charge carrier mobility. 

\begin{figure*}
\includegraphics[width=\textwidth]{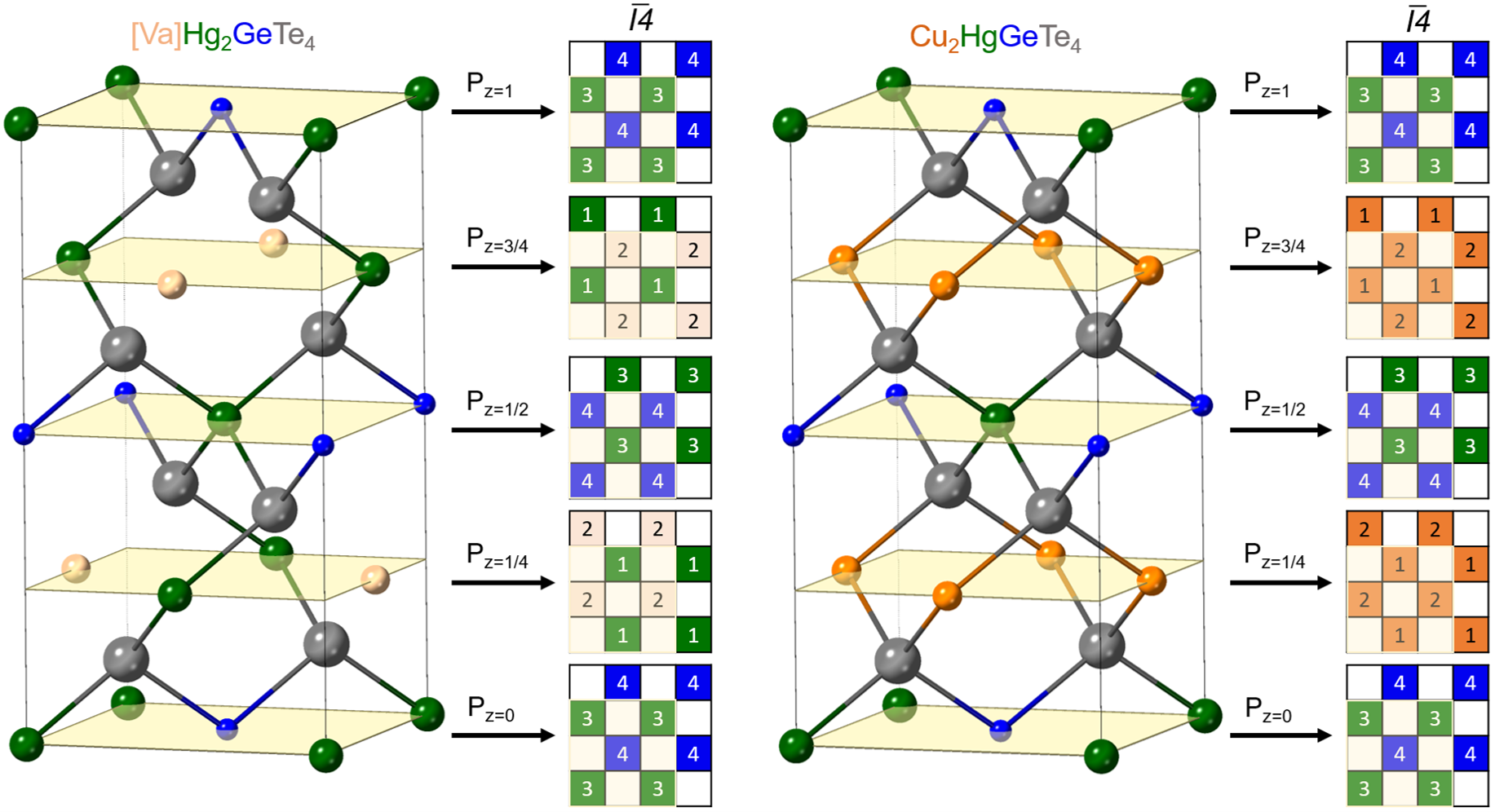}
\caption{\label{schematic}The ternary endpoint, $\mathrm{[Va]Hg_2GeTe_4}$, of the $\mathrm{Cu_{2x}Hg_{2-x}GeTe_4}$  alloy crystallizes in the defect chalcopyrite structure and the quaternary endpoint, $\mathrm{Cu_2HgGeTe_4}$, crystallizes in the stannite structure. The unit cell for these two compounds is shown here using the ${I\overline{4}}$ space group. The location of specific atomic sites, as defined by the ${I\overline{4}}$ space group, are denoted numerically for each endpoint on the z = 0, 1/4, 1/2, 3/4 and 1 planes and are colored based on the atom occupying that site for each composition---where vacancies are beige, Cu atoms are orange, Hg atoms are green, Ge atoms are blue and Te atoms are gray. Each schematic plane represents a 2x2 unit cell, which is used later in DFT simulations, and the atoms contained in the original unit cell are highlighted within the yellow boxes on each plane.}
\end{figure*}

One class of thermoelectric materials that demonstrates promise for achieving low thermal conductivity is the quaternary diamond-like semiconductors (DLS), such as $\mathrm{Cu_2(IIB)(IV)Te_4}$ (IIB: Zn, Cd, Hg) (IV: Si, Ge, Sn) \cite{Brenden1}. Among these compounds, $\mathrm{Cu_2HgGeTe_4}$ was shown to exhibit unusually low thermal conductivity  ($\kappa_L <$ 0.25 W/mK) and high hole mobility ($\mu_0>50$ $\mathrm{cm^2/Vs}$) \cite{Brenden1}. The low thermal conductivity of $\mathrm{Cu_2HgGeTe_4}$ was attributed to strong phonon scattering from $\mathrm{Cu_{Hg}}$ and $\mathrm{Hg_{Cu}}$ anti-site defects, but further improvements in the overall efficiency of $\mathrm{Cu_2HgGeTe_4}$  requires an optimization of the carrier concentration.

Controlling carrier concentration in the DLS family of materials via manipulation of defects and dopants has been the topic of a number of previous studies \cite{doi:10.1002/aenm.201502386, doi:10.1063/1.4872250, Kosuga_2012, doi:10.1063/1.3678044, doi:10.1063/1.3617458, doi:10.1002/aenm.201601299, doi:10.1063/1.4902849, ZHANG2017156, Zhou:2017:1941-4900:1520, CHETTY201617, doi:10.1002/adma.200900409, doi:10.1063/1.3103604}. Our previous work revealed the existence of a full solid solution between $\mathrm{Cu_2HgGeTe_4}$ and $\mathrm{Hg_2GeTe_4}$, denoted here as $\mathrm{Cu_{2x}Hg_{2-x}GeTe_4}$ (where 0 $\leq$ x $\leq$ 1) \cite{C8TA10332A}. We demonstrated that the extent of Cu integration in $\mathrm{Cu_{2x}Hg_{2-x}GeTe_4}$ could be used to manipulate the carrier concentration, from degenerate ($\mathrm{>10^{21}}$ $\mathrm{h^+}$ $\mathrm{cm^{-3}}$) in $\mathrm{Cu_2HgGeTe_4}$ to intrinsic ($\mathrm{<10^{17}}$ $\mathrm{h^+}$ $\mathrm{cm^{-3}}$) in $\mathrm{Hg_2GeTe_4}$ \cite{C8TA10332A}. Considering that the alloy remained charge balanced, the mechanism for varying the carrier concentration was unknown.

The structure of the endpoints of the  $\mathrm{Cu_{2x}Hg_{2-x}GeTe_4}$  alloy are known to be stannite (${I\overline{4}2m}$) for $\mathrm{Cu_2HgGeTe_4}$ \cite{Brenden1} and defect chalcopyrite (${I\overline{4}}$) for $\mathrm{Hg_2GeTe_4}$ \cite{C8TA10332A}. Since the ${I\overline{4}2m}$ space group is a maximal subgroup of the ${I\overline{4}}$ space group, ${I\overline{4}}$ can be used to describe the structure of both $\mathrm{Cu_2HgGeTe_4}$ and $\mathrm{Hg_2GeTe_4}$. The unit cell for these two endpoints of the $\mathrm{Cu_{2x}Hg_{2-x}GeTe_4}$  alloy are shown in Fig. \ref{schematic} along with a schematic that shows the location of each atomic site on the z = 0, 1/4, 1/2, 3/4, 1 planes (see SI Fig. 1 for unit cells of all compositions).  For the sake of clarity, $\mathrm{Hg_2GeTe_4}$ is denoted as $\mathrm{[Va]Hg_2GeTe_4}$ in Fig. \ref{schematic} to emphasize the vacancy on site 2 for that structure.   

\begin{figure*}
\includegraphics[width=0.53\textwidth]{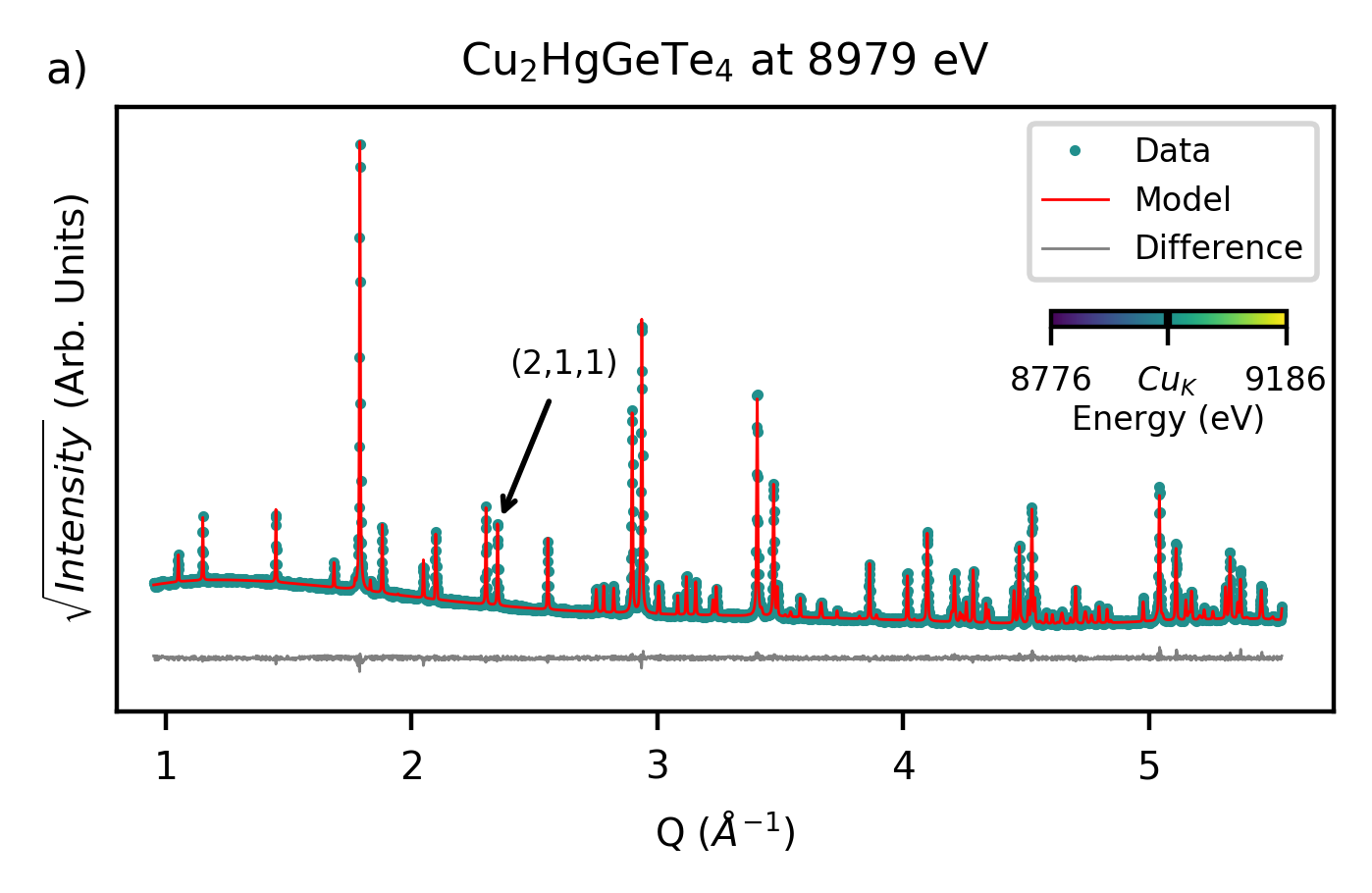}
\includegraphics[width=0.47\textwidth]{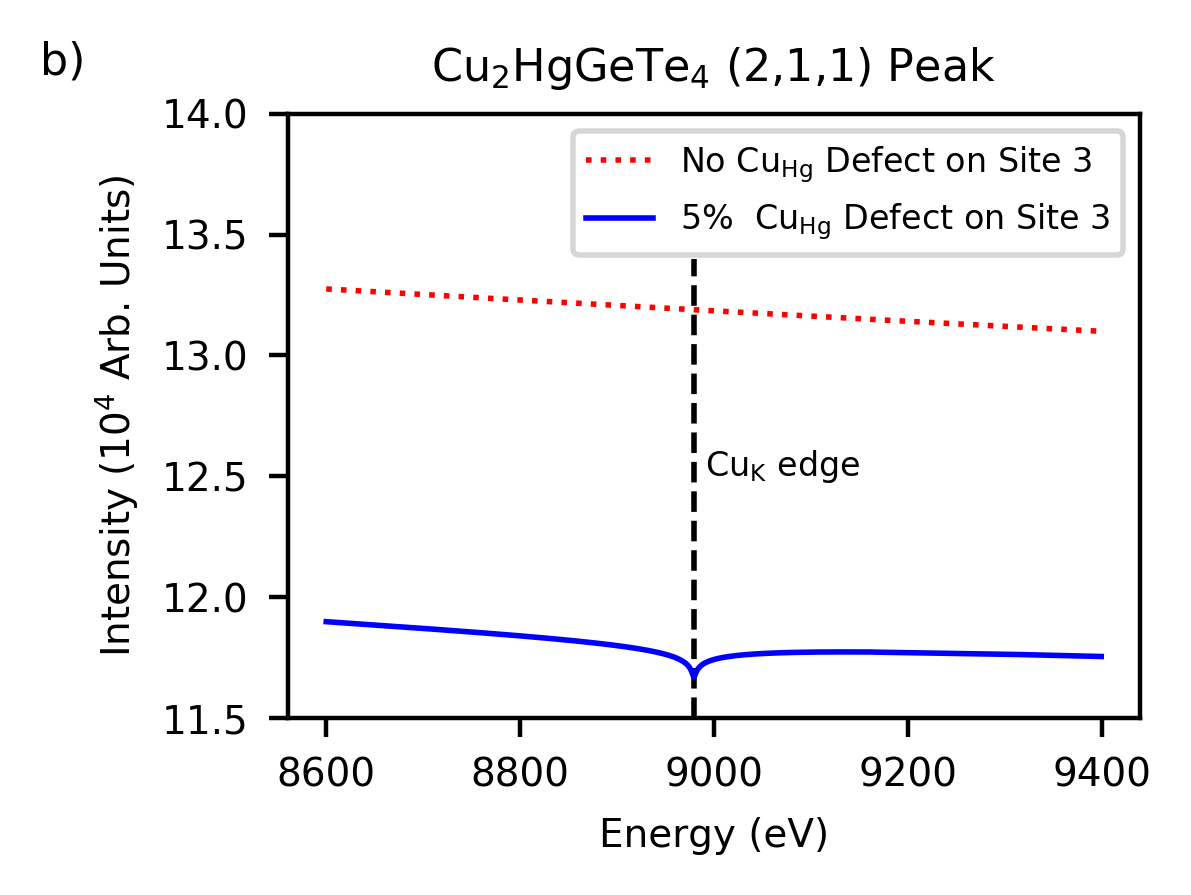}
\caption{\label{REXD_method} a) A sample REXD pattern collected for the sample at the $\mathrm{Cu_k}$ edge (8979 eV). The (2,1,1) peak is pointed out here with an arrow. b) REXD simulations of the (2,1,1) peak for the quaternary sample demonstrating the difference between a sample with no anti-site defects and a sample with 5\% $\mathrm{Cu_{Hg}}$ anti-site defects. The sample with 5\% anti-site defects demonstrates a sudden decrease in intensity across the $\mathrm{Cu_k}$ edge that is not present for the sample without anti-site defects.}
\end{figure*}

In this work, we use a combination of REXD and density functional theory (DFT) calculations to  characterize the $\mathrm{Cu_{2x}Hg_{2-x}GeTe_4}$ crystal structure as a function of alloy composition. We find that vacancy ordering, where the vacancies prefer to occupy a single site, is maintained across all alloy compositions and that Cu incorporates into the alloy structure by preferentially occupying sites 1 and 2 of the z = 1/4 plane equally before moving on to the z = 3/4 plane (see Fig. \ref{schematic}). Furthermore, we show that the extent of $\mathrm{Cu_{Hg}}$ anti-site defects increases in direct proportionality with the experimentally measured hole concentrations. This work demonstrates how in-depth structural characterizations, including quantification of point defects, can provide valuable insight into how a material's structure affects electronic properties.

\section{Materials and Methods}
\subsubsection{Experimental}
Powder samples of $\mathrm{Cu_{2x}Hg_{2-x}GeTe_4}$ (x=0, 0.2, 0.4, 0.6, 0.8, 1) from our previous study were also used here. We synthesized these  via solid-state reaction as described previously \cite{C8TA10332A}. All samples contain a trace amount of intentional impurities from the phase boundary mapping process \cite{C8TA10332A} ( $< 1.80\ \%$ of HgTe and $< 5.02\ \%$ of GeTe) as determined by our XRD phase fraction analysis.

High resolution X-ray diffraction (HRXRD) experiments were conducted at beamline 11-BM of the Advanced Photon Source (APS) with a wavelength of 0.413 \AA. Data was collected by rotating the detector array of 12 independent point detectors (spaced apart by $\sim 2^\circ$) from $2^\circ$ – $28^\circ$, thus covering an angular range from $2^\circ$ – $50^\circ$, in increments of $0.001^\circ$ at a scan speed of $0.01^\circ/s$. All samples were diluted with a 1:10 mol ratio of amorphous $\mathrm{SiO_2}$ before being loaded into 0.5 mm glass capillaries (special glass, Charles Supper).

Resonant X-Ray Diffraction measurements were carried out at both beamline 2-1 of the Stanford Synchrotron Radiation Lightsource (SSRL) and beamline 33-BM of the APS. In both cases, samples were measured under Bragg-Brentano geometry using Si zero background plates and scattered X-rays were detected using a Pilatus 100K area detector. All samples were measured under an inert gas environment. Full powder diffraction patterns were measured across the $\mathrm{Cu_k}$ absorption edge for every sample. $\mathrm{Cu_{0.8}Hg_{1.6}GeTe_4}$ was measured at 33-BM and all remaining Cu containing samples ($\mathrm{Cu_{2x}Hg_{2-x}GeTe_4}$ where x = 0.2, 0.6, 0.8, 1) were measured at beamline 2-1. Samples measured at 2-1 were rocked by $\pm0.5 ^\circ$C during measurements to enhance powder averaging, but samples measured at 33-BM were not rocked.

\begin{figure*}
\includegraphics[width=\textwidth]{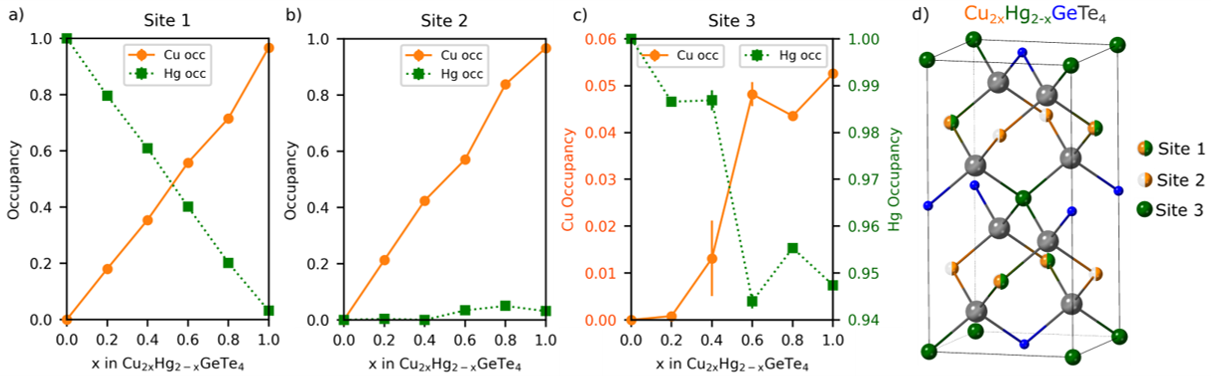}
\caption{\label{occupancies} a-c) The Cu and Hg occupancies are plotted here for sites 1, 2 and 3. All occupancies were obtained via Rietveld refinement of REXD data on the $\mathrm{Cu_{2x}Hg_{2-x}GeTe_4}$ alloys at room temperature. d) An example unit cell for the $\mathrm{Cu_{2x}Hg_{2-x}GeTe_4}$ alloy composition is shown here with sites 1, 2 and 3 indicated by the legend. }
\end{figure*}

All synchrotron diffraction data was analyzed using the TOPAS Academic software package and modelled to a unified structural model using Rietveld refinement. This structural model was based upon the lower symmetry $\mathrm{Hg_2GeTe_4}$ compound, which has space group ${I\overline{4}}$. The lower degree of symmetry of this space group has the advantage that it enables the use of a single structural model for all measured alloy compositions.

The first step in the analysis process was to perform Rietveld refinement on the HRXRD data. For each sample composition, the data was analyzed by refinement on lattice parameters, atomic occupancies, thermal parameters and atomic position of the Te atom in site 5. While the sites 1, 2, 3 and 4 atoms all lie on special positions within the unit cell, the atomic position of the Te atom in site 5 was refined as (x, y, z). Peak broadening was fit using a combination of Gaussian and Lorentzian contributions from size and microstrain parameters. Errors were obtained using the bootstrapping method with 25 iterations \cite{efron1986, diciccio1996, chernick_2007}.

The results of the HRXRD refinement (see SI Fig. 2-7) were then used as the starting point for the REXD refinement. For a given sample composition, all the REXD patterns collected at different energies were co-refined using a single structural model (see SI Ex. 1 for a sample input file). All non-occupancy parameters (i.e., thermal parameters, lattice parameters and Te atom position) were fixed to the values determined by HRXRD because the HRXRD scans contain data at a higher Q range, which is most important for refining these parameters. This allows us to focus on refining only the atomic occupancies in the REXD refinements.  Peak broadening was again fit using a combination of Gaussian and Lorentzian contributions from size and strain parameters, but further corrections were needed due to anisotropic peak broadening. For this correction, we used Stephen’s tetragonal model \cite{Stephens:hn0085}. The REXD refinement also required the use of a surface roughness correction to account for over-compaction of the powders in certain measurements (x = 0.2, 0.6, 0.8). For this correction, we used a macro developed by Suortti \cite{Suortti:a09574}. The total occupancy for each site was constrained such that it could not exceed one and quadratic penalties were used to constrain the overall composition to the nominal stoichiometric value. 

An example of the final refinements is shown in Fig. \ref{REXD_method}a for the $\mathrm{Cu_2HgGeTe_4}$ sample at the $\mathrm{Cu_k}$ edge (see SI GIFs 1-5 for animations of the final REXD refinements of each composition as a function of energy). The (2,1,1) peak is pointed out because it demonstrates the highest sensitivity to the presence of $\mathrm{Cu_{Hg}}$defects, as indicated by the sudden drop in intensity across the $\mathrm{Cu_k}$ edge in Fig. \ref{REXD_method}b. While modest, this change in intensity enables reliable quantification of anti-site defects in our REXD refinements. 

\subsubsection{Computational}
Investigation of the energetics of the alloys was performed using first-principles calculations within the DFT formalism~\cite{hohenberg-kohn-1964,kohn-sham-1965}. The DFT calculations were performed with the plane-wave basis Vienna Ab initio Simulation Package (VASP) \cite{kressePRB1996}. The generalized gradient approximation (GGA) of Perdew-Burke-Ernzerhof (PBE) \cite{Perdew1996} in the projector augmented wave formalism \cite{blochlPRB1994} was used. The Kohn-Sham orbitals were expanded using a plane-wave basis with a cutoff energy of 500 eV. 

Total energies of $\mathrm{Hg_2GeTe_4}$ and $\mathrm{Cu_2HgGeTe_4}$ were computed using conventional cells composed of 14 and 16 atoms respectively. In these cases, the Brillouin zone was sampled using a $\Gamma$-centered 8x8x4 Monkhorst-Pack k-point grid \cite{mp-kgrid}. 

\begin{figure}[t]
\includegraphics[width=\columnwidth]{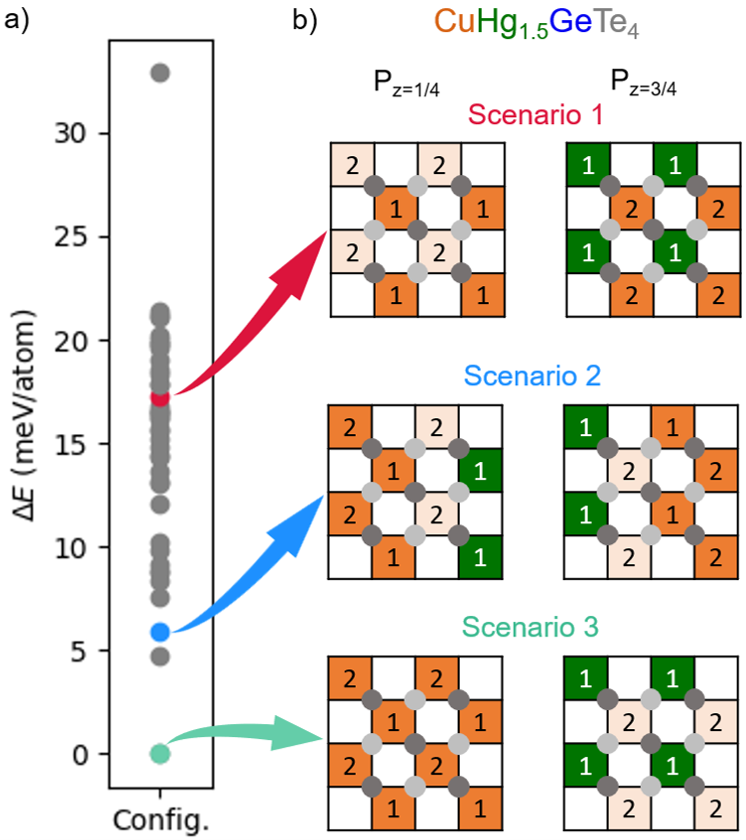}
\caption{\label{dft_fig} a) DFT total energies for approximately 50 permutations (gray dots) of the intermediate $\mathrm{CuHg_{1.5}GeTe_4}$ alloy. The energies are referenced to the lowest energy configuration, shown by the light-green dot. b) The lowest energy permutation of each of the three possible scenarios from REXD results is demonstrated schematically. Scenario 3, where Cu occupies sites 1 and 2 exclusively on the z = 1/4 plane, has the lowest energy and is therefore the most favorable. Te atoms that sit above and below the z = 1/4 and 3/4 planes are shown in light gray and dark gray circles respectively.}
\end{figure}

The energy cost of the defects in the ternary, quaternary and intermediate composition compounds were evaluated using a 2x2x1 supercell (as shown schematically in Fig. \ref{schematic}) and a $\Gamma$-centered 4x4x4 Monkhorst-Pack k-point grid. For the systems with intermediate composition, around 50 supercells with different site occupations were used. In this case, the supercells were generated with the Clusters Approach to Statistical Mechanics (CASM) open-source software package \cite{CASM}.  

\section{Results and Discussion}
To understand how Cu integrates into the $\mathrm{Cu_{2x}Hg_{2-x}GeTe_4}$ structure, we use REXD across the $\mathrm{Cu_K}$ edge to probe the atomic occupancies of sites 1, 2 and 3 for each Cu-containing compound (x = 0.2, 0.4, 0.6, 0.8 and 1). HRXRD was used for $\mathrm{Hg_2GeTe_4}$ (x = 0) since this compound contains no Cu. These occupancies, plotted in Fig. \ref{occupancies}a-c, provide insight into how much of each type of atom lies on a specific lattice site (see SI Tab. 1 for a table of REXD results). As indicated in Fig. \ref{occupancies}d, site 1 refers to the site that is nominally occupied by some combination of Hg and Cu depending on the sample composition, site 2 refers to the site that is vacant in $\mathrm{Hg_2GeTe_4}$ and nominally occupied by Cu in $\mathrm{Cu_2HgGeTe_4}$, and site 3 is nominally occupied by Hg for all compositions. REXD across the $\mathrm{Cu_K}$ edge enables more accurate quantification of Cu atomic occupancies than would be possible using non-resonant XRD due to the enhanced sensitivity from the Cu resonance effect. This enhanced sensitivity to Cu is particularly critical for probing $\mathrm{Cu_{Hg}}$ anti-site defects on site 3.
 
From Fig. \ref{occupancies}a-b at x = 0, it is apparent that site 1 is initially fully occupied by Hg and site 2 is completely vacant, indicating that there is ordering of the vacancies and Hg atoms between sites 1 and 2 on the z = 1/4 and z = 3/4 planes in $\mathrm{Hg_2GeTe_4}$. If Hg exists in $\mathrm{Hg_2GeTe_4}$ in the $\mathrm{Hg^{2+}}$ state and Cu integrates into the alloy as $\mathrm{Cu^{1+}}$, then it follows that charge balance is maintained by exchanging Cu and Hg in a 2:1 ratio as the alloy progresses from $\mathrm{Hg_2GeTe_4}$ (x = 0) to $\mathrm{Cu_2HgGeTe_4}$ (x = 1).

Examining the occupancies of sites 1 and 2 in Fig. \ref{occupancies}a and \ref{occupancies}b respectively reveals that Cu fills these two sites roughly equally as the alloy composition changes from x = 0 to x = 1. Furthermore, the amount of additional Cu introduced into the alloy in sites 1 and 2 at each sequential composition is consistently twice the amount of Hg that is removed from site 1. Taken together, these results indicate that Cu fills the $\mathrm{Cu_{2x}Hg_{2-x}GeTe_4}$ lattice in a 2:1 ratio with Hg wherein one Cu atom annihilates a Hg vacancy on site 2 while another Cu atom swaps with an existing Hg atom on site 1.

Meanwhile, Hg continues to prefer occupying site 1 rather than site 2 for all compositions --- as evidenced by significantly higher occupancies of Hg on site 1 in Fig. \ref{occupancies}a than on site 2 in Fig. \ref{occupancies}b --- except for $\mathrm{Cu_2HgGeTe_4}$, where there is very little Hg on either site.  However, there is a slight increase in the Hg occupancy of site 2 above the vacancy levels for the x = 0.6, 0.8 and 1 compositions in Fig. \ref{occupancies}b, which points to the possibility of $\mathrm{Cu_{Hg}}$ anti-site defects in the alloy system. 

To further understand the possible source of these defects, the occupancy of site 3 is plotted in Fig. \ref{occupancies}c where we see that the Hg occupancy decreases and is accompanied by a simultaneous increase in the Cu site 3 occupancy as the alloy approaches the x = 1 composition. The presence of Cu on site 3 demonstrates the presence of $\mathrm{Cu_{Hg}}$ anti-site defects. The amount of these defects increases as the alloy composition moves closer to x = 1 ($\mathrm{Cu_2HgGeTe_4}$). The propensity for $\mathrm{Cu_2HgGeTe_4}$ to demonstrate $\mathrm{Cu_{Hg}}$ swaps and for $\mathrm{Hg_2GeTe_4}$ to be strongly ordered (i.e., no $\mathrm{Hg_{Va}}$ swaps) is confirmed by DFT calculations, which reveal that the energy cost $\mathrm{\Delta E}$ for $\mathrm{Cu_{Hg}}$ swaps in  $\mathrm{Cu_2HgGeTe_4}$ ($\mathrm{\Delta E}$ = 0.23 eV) is about three times smaller than for $\mathrm{Hg_{Va}}$ swaps in $\mathrm{Hg_2GeTe_4}$ ($\mathrm{\Delta E}$ = 0.67 - 0.76 eV).

In brief summary, REXD experiments have revealed that Cu integrates into the $\mathrm{Cu_{2x}Hg_{2-x}GeTe_4}$ alloy structure in a 2:1 ratio with Hg where Cu simultaneously  swaps with a Hg atom on site 1 and annihilates a vacancy on site 2. We also know that Hg-vacancy ordering is strongest at compositions closest to x = 0 ($\mathrm{Hg_2GeTe_4}$) and that there is an increasing likelihood of $\mathrm{Cu_{Hg}}$ anti-site defects on site 3 as Cu incorporates into the alloy composition. However, REXD experiments are still unable to tell us whether the Cu integration occurs primarily on one of the z = 1/4 or z = 3/4 planes or on both simultaneously. This leaves three possible scenarios for how Cu integrates into the $\mathrm{Cu_{2x}Hg_{2-x}GeTe_4}$ structure. These three scenarios are illustrated in Fig. \ref{dft_fig}b for the intermediate x = 0.5 alloy composition ($\mathrm{CuHg_{1.5}GeTe_4}$) and are described as follows: Scenario 1) Cu occupies site 1 on one plane and site 2 on the other plane, Scenario 2) Cu occupies sites 1 and 2 on both planes equally, Scenario 3) Cu occupies sites 1 and 2 exclusively on one plane. Due to site symmetry, we are unable to differentiate between these three possible scenarios using REXD, so we turn to DFT calculations performed on the intermediate composition (x = 0.5) to see what the preferred structure is on the basis of total energy. Fig. \ref{dft_fig}a shows the total energy in meV/atom for approximately 50 potential configurations of the x = 0.5 alloy. The three configurations highlighted in green, blue and red show the lowest energy permutation of each of the three scenarios described previously. Of these, scenario 3 has the lowest energy, indicating that Cu prefers to fill one plane entirely before beginning to fill the other plane as Cu is incorporated into the $\mathrm{Cu_{2x}Hg_{2-x}GeTe_4}$ structure. 

One possible explanation for why scenario 3 is lower in energy than the other scenarios is based on electrostatics. In Fig. \ref{dft_fig}b, the Te atoms that sit above and below the z = 1/4 and z = 3/4 planes are shown in light gray and dark gray circles respectively. Looking at the $\mathrm{Hg_2GeTe_4}$ and $\mathrm{Cu_2HgGeTe_4}$ unit cells in Fig. \ref{schematic}, we see that each Te atom is bonded to one Ge atom and at least one Hg atom (for instance, from the z = 0 or z = 1/2 planes) regardless of composition. Assuming Ge has an oxidation state of $\mathrm{4^+}$ and Hg has an oxidation state of $\mathrm{2^+}$, then each Te atom needs to be surrounded by an additional charge of $\mathrm{2^+}$ (in the z = 1/4 and 3/4 planes) in order for the octet rule to be satisfied. This is accomplished in $\mathrm{Cu_2HgGeTe_4}$ by two $\mathrm{Cu^{1+}}$ atoms and in $\mathrm{Hg_2GeTe_4}$ by one $\mathrm{Hg^{2+}}$ atom and a vacancy. Similarly, for the intermediate compositions in Fig. \ref{dft_fig}b, the octet rule will be satisfied whenever the gray dots representing Te atoms are between two atoms, or boxes, for which the net charge is $\mathrm{2^+}$. Since Cu (orange box) has an oxidation state of $\mathrm{1^+}$, Hg (green box) has an oxidation state of $\mathrm{2^+}$ and vacancies (beige box) have no charge, Scenario 3 is the only scenario that satisfies the octet rule for all Te atoms. This argument is supported by the observation that in Scenario 1 none of the Te atoms satisfy the octet rule and Scenario 1 has the highest energy. 

\begin{figure}[t]
\includegraphics[width=\columnwidth]{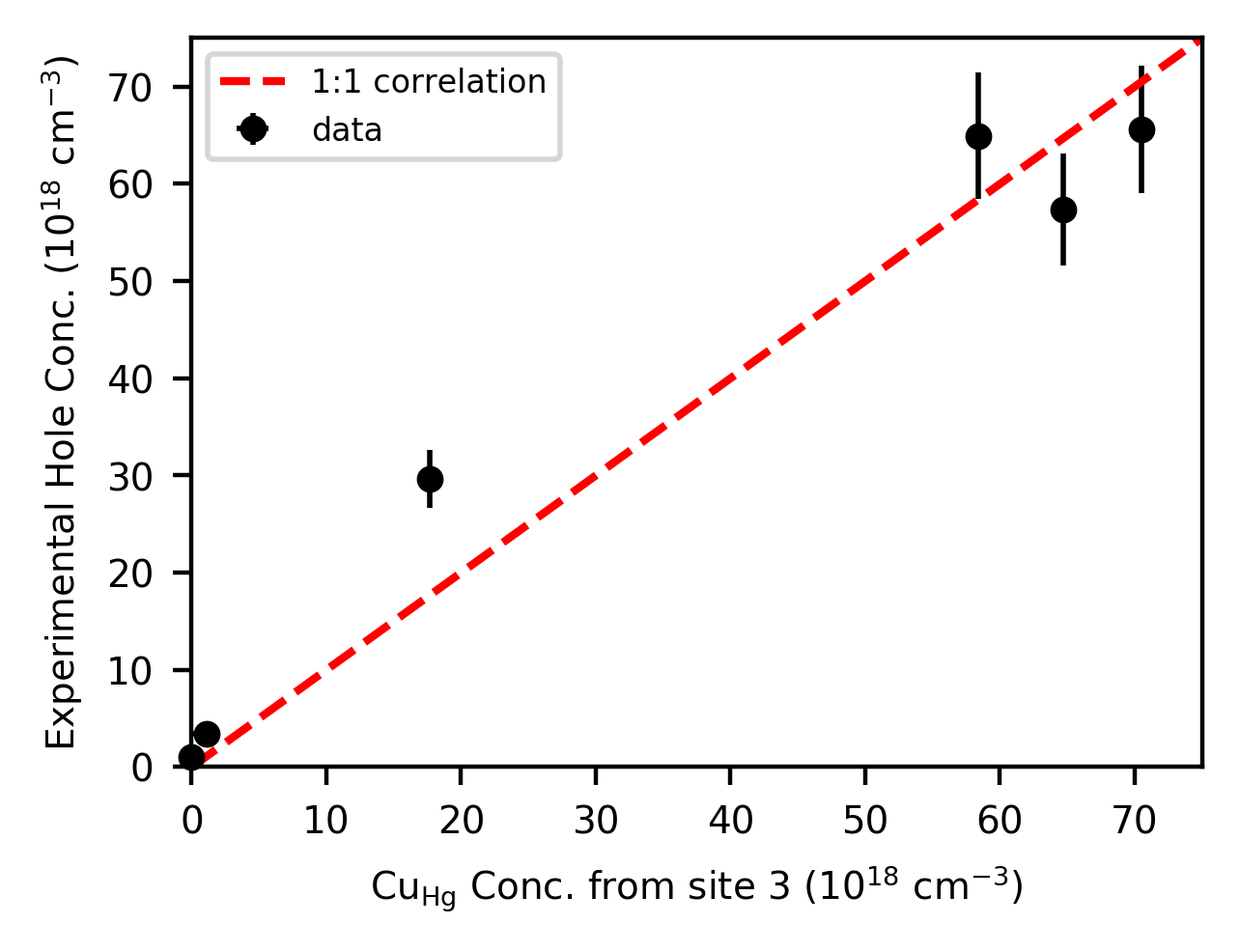}
\caption{\label{props_fig} The experimentally measured hole concentration, taken from \cite{C8TA10332A} (with a standard error of 10\%), is plotted here as a function of the site 3 $\mathrm{Cu_{Hg}}$ concentration from the current study. A line with a slope of 1 is also drawn to illustrate the strong correlation, which indicates that $\mathrm{Cu_{Hg}}$ anti-site defects are responsible for controlling the carrier concentration of $\mathrm{Cu_{2x}Hg_{2-x}GeTe_4}$ alloys.}
\end{figure}
The discussion up to this point has provided a detailed understanding of how Cu is integrated into the $\mathrm{Cu_{2x}Hg_{2-x}GeTe_4}$ structure, so we now turn our attention to how this structure affects the thermoelectric properties. Converting the site 3 Cu occupancies from Fig. \ref{occupancies} into $\mathrm{Cu_{Hg}}$ concentrations (as explained in SI Eq. 1) revealed a direct proportionality with previously measured hole concentrations \cite{C8TA10332A} across the alloy composition range, as shown in Fig. \ref{props_fig}. This suggests that $\mathrm{Cu_{Hg}}$ anti-site defects, which are the source of Cu on site 3,  are responsible for tuning the carrier concentration in the $\mathrm{Cu_{2x}Hg_{2-x}GeTe_4}$ alloys.

The ability to tune carrier concentration via manipulation of anti-site defects is a striking result worth further consideration. While it is intuitive that $\mathrm{Cu_{Hg}}$ anti-site defects might increase as the alloy composition becomes more Cu rich, it was unexpected to see that the correlation between the site 3 Cu concentration and the experimentally measured hole concentrations was directly proportional (e.g., 1:1) because this implies that no other defects contribute to the carrier concentration. In particular, we expected that Cu vacancies ($\mathrm{V_{Cu}}$) may also contribute to the carrier concentration via hole generation. However, two key facts justify our implicit disregard for ($\mathrm{V_{Cu}}$) here. First is that DFT defect diagrams for the Cu-rich $\mathrm{Cu_{2}HgGeTe_4}$ compound predict $\mathrm{Cu_{Hg}}$ as the dominant defect \cite{Jiaxing}. Second is that if $\mathrm{V_{Cu}}$ was responsible for the increased carrier concentration in the $\mathrm{Cu_{2x}Hg_{2-x}GeTe_4}$ alloy, then we would expect to see a decrease in the carrier concentration as excess Cu is incorporated into $\mathrm{Cu_{2}HgGeTe_4}$ because this excess Cu would annihilate $\mathrm{V_{Cu}}$. However, our previous work demonstrated that integrating excess Cu into $\mathrm{Cu_{2}HgGeTe_4}$ actually led to a higher hole concentration \cite{C8TA10332A}, indicating that $\mathrm{Cu_{Hg}}$ --- not $\mathrm{V_{Cu}}$ --- is the predominant defect in the alloy system. This work therefore demonstrates that $\mathrm{Cu_{Hg}}$ is a favorable defect and singularly determines the carrier concentration along the entire alloying series. Thus, these results justify our previous observations and demonstrate how REXD can provide immense insight into the defects of complex materials.

\section{Conclusion}
We have shown through a combination of REXD experiments and DFT calculations that Cu integrates into the $\mathrm{Cu_{2x}Hg_{2-x}GeTe_4}$ alloy compound in a 2:1 ratio with Hg where Cu simultaneously annihilates a vacancy and swaps with a Hg atom on the z = 1/4 and z = 3/4 planes.  Furthermore, we showed that the ordering of vacancies is maintained as Cu is incorporated into the structure and that Cu atoms prefer to fill a single plane entirely before moving on to the next one. The presence of $\mathrm{Cu_{Hg}}$ anti-site defects was quantified by REXD and shown to directly control the measured hole concentrations. These results indicate a systematic and ordered incorporation of Cu into $\mathrm{Cu_{2x}Hg_{2-x}GeTe_4}$ rather than a randomized entropic process, which explains the linear nature of the effect of Cu incorporation on carrier concentration. The ability of REXD to both quantify the presence of anti-site defects and understand how these defects affect the carrier concentration can have a significant impact on the development of future thermoelectric materials and semiconductors.
 
\begin{acknowledgments}
B.L.W, B.R.O, E.S.T and M.F.T acknowledge support from the National Science Foundation, DMREF No. 1729594. L.C.G. and E.E acknowledge support from the National Science Foundation, DMREF No. 1729149. B.L.W. acknowledges support from the National Science Foundation Graduate Research Fellowship under Grant No. DGE-114747.  Use of the Stanford Synchrotron Radiation Lightsource, SLAC National Accelerator Laboratory, is supported by the DOE Office of Science (SC), Basic Energy Sciences (BES) under Contract No. DE-AC02-76SF00515. Use of the Advanced Photon Source was supported by the U. S. Department of Energy, Office of Science, Office of Basic Energy Sciences, under Contract No. DE-AC02-06CH11357. We thank Saul Lapidus and Jenia Karapetrova for their support at APS beamlines 11-BM and 33-BM  respectively. Computational resources were provided by the Blue Waters sustained-petascale computing project, which is supported by the National Science Foundation (awards OCI-0725070 and ACI-1238993) the State of Illinois, and as of December, 2019, the National Geospatial-Intelligence Agency. Blue Waters is a joint effort of the University of Illinois at Urbana-Champaign and its National Center for Supercomputing Applications.
\end{acknowledgments}

\bibliography{PRMaterials_paper}

\providecommand{\noopsort}[1]{}\providecommand{\singleletter}[1]{#1}%
\begin{thebibliography}{34}%
\makeatletter
\providecommand \@ifxundefined [1]{%
 \@ifx{#1\undefined}
}%
\providecommand \@ifnum [1]{%
 \ifnum #1\expandafter \@firstoftwo
 \else \expandafter \@secondoftwo
 \fi
}%
\providecommand \@ifx [1]{%
 \ifx #1\expandafter \@firstoftwo
 \else \expandafter \@secondoftwo
 \fi
}%
\providecommand \natexlab [1]{#1}%
\providecommand \enquote  [1]{``#1''}%
\providecommand \bibnamefont  [1]{#1}%
\providecommand \bibfnamefont [1]{#1}%
\providecommand \citenamefont [1]{#1}%
\providecommand \href@noop [0]{\@secondoftwo}%
\providecommand \href [0]{\begingroup \@sanitize@url \@href}%
\providecommand \@href[1]{\@@startlink{#1}\@@href}%
\providecommand \@@href[1]{\endgroup#1\@@endlink}%
\providecommand \@sanitize@url [0]{\catcode `\\12\catcode `\$12\catcode
  `\&12\catcode `\#12\catcode `\^12\catcode `\_12\catcode `\%12\relax}%
\providecommand \@@startlink[1]{}%
\providecommand \@@endlink[0]{}%
\providecommand \url  [0]{\begingroup\@sanitize@url \@url }%
\providecommand \@url [1]{\endgroup\@href {#1}{\urlprefix }}%
\providecommand \urlprefix  [0]{URL }%
\providecommand \Eprint [0]{\href }%
\providecommand \doibase [0]{https://doi.org/}%
\providecommand \selectlanguage [0]{\@gobble}%
\providecommand \bibinfo  [0]{\@secondoftwo}%
\providecommand \bibfield  [0]{\@secondoftwo}%
\providecommand \translation [1]{[#1]}%
\providecommand \BibitemOpen [0]{}%
\providecommand \bibitemStop [0]{}%
\providecommand \bibitemNoStop [0]{.\EOS\space}%
\providecommand \EOS [0]{\spacefactor3000\relax}%
\providecommand \BibitemShut  [1]{\csname bibitem#1\endcsname}%
\let\auto@bib@innerbib\@empty
\bibitem [{\citenamefont {Schnepf}\ \emph
  {et~al.}(2020{\natexlab{a}})\citenamefont {Schnepf}, \citenamefont {Cordell},
  \citenamefont {Tellekamp}, \citenamefont {Melamed}, \citenamefont
  {Greenaway}, \citenamefont {Mis}, \citenamefont {Brennecka}, \citenamefont
  {Christensen}, \citenamefont {Tucker}, \citenamefont {Toberer}, \citenamefont
  {Lany},\ and\ \citenamefont {Tamboli}}]{doi:10.1021/acsenergylett.0c00576}%
  \BibitemOpen
  \bibfield  {author} {\bibinfo {author} {\bibfnamefont {R.~R.}\ \bibnamefont
  {Schnepf}}, \bibinfo {author} {\bibfnamefont {J.~J.}\ \bibnamefont
  {Cordell}}, \bibinfo {author} {\bibfnamefont {M.~B.}\ \bibnamefont
  {Tellekamp}}, \bibinfo {author} {\bibfnamefont {C.~L.}\ \bibnamefont
  {Melamed}}, \bibinfo {author} {\bibfnamefont {A.~L.}\ \bibnamefont
  {Greenaway}}, \bibinfo {author} {\bibfnamefont {A.}~\bibnamefont {Mis}},
  \bibinfo {author} {\bibfnamefont {G.~L.}\ \bibnamefont {Brennecka}}, \bibinfo
  {author} {\bibfnamefont {S.}~\bibnamefont {Christensen}}, \bibinfo {author}
  {\bibfnamefont {G.~J.}\ \bibnamefont {Tucker}}, \bibinfo {author}
  {\bibfnamefont {E.~S.}\ \bibnamefont {Toberer}}, \bibinfo {author}
  {\bibfnamefont {S.}~\bibnamefont {Lany}},\ and\ \bibinfo {author}
  {\bibfnamefont {A.~C.}\ \bibnamefont {Tamboli}},\ }\bibfield  {title}
  {\bibinfo {title} {Utilizing site disorder in the development of new
  energy-relevant semiconductors},\ }\href
  {https://doi.org/10.1021/acsenergylett.0c00576} {\bibfield  {journal}
  {\bibinfo  {journal} {ACS Energy Letters}\ }\textbf {\bibinfo {volume} {5}},\
  \bibinfo {pages} {2027} (\bibinfo {year} {2020}{\natexlab{a}})}\BibitemShut
  {NoStop}%
\bibitem [{\citenamefont {Pan}\ \emph {et~al.}(2020)\citenamefont {Pan},
  \citenamefont {Cordell}, \citenamefont {Tucker}, \citenamefont {Zakutayev},
  \citenamefont {Tamboli},\ and\ \citenamefont {Lany}}]{Pan2020}%
  \BibitemOpen
  \bibfield  {author} {\bibinfo {author} {\bibfnamefont {J.}~\bibnamefont
  {Pan}}, \bibinfo {author} {\bibfnamefont {J.~J.}\ \bibnamefont {Cordell}},
  \bibinfo {author} {\bibfnamefont {G.~J.}\ \bibnamefont {Tucker}}, \bibinfo
  {author} {\bibfnamefont {A.}~\bibnamefont {Zakutayev}}, \bibinfo {author}
  {\bibfnamefont {A.~C.}\ \bibnamefont {Tamboli}},\ and\ \bibinfo {author}
  {\bibfnamefont {S.}~\bibnamefont {Lany}},\ }\bibfield  {title} {\bibinfo
  {title} {Perfect short-range ordered alloy with line-compound-like properties
  in the $\mathrm{ZnSnN_2:ZnO}$ system},\ }\href
  {https://doi.org/10.1038/s41524-020-0331-8} {\bibfield  {journal} {\bibinfo
  {journal} {NPJ Computational Materials}\ }\textbf {\bibinfo {volume} {6}},\
  \bibinfo {pages} {63} (\bibinfo {year} {2020})}\BibitemShut {NoStop}%
\bibitem [{\citenamefont {Schnepf}\ \emph
  {et~al.}(2020{\natexlab{b}})\citenamefont {Schnepf}, \citenamefont
  {Levy-Wendt}, \citenamefont {Tellekamp}, \citenamefont {Ortiz}, \citenamefont
  {Melamed}, \citenamefont {Schelhas}, \citenamefont {Stone}, \citenamefont
  {Toney}, \citenamefont {Toberer},\ and\ \citenamefont {Tamboli}}]{Rekha}%
  \BibitemOpen
  \bibfield  {author} {\bibinfo {author} {\bibfnamefont {R.~R.}\ \bibnamefont
  {Schnepf}}, \bibinfo {author} {\bibfnamefont {B.~L.}\ \bibnamefont
  {Levy-Wendt}}, \bibinfo {author} {\bibfnamefont {M.~B.}\ \bibnamefont
  {Tellekamp}}, \bibinfo {author} {\bibfnamefont {B.~R.}\ \bibnamefont
  {Ortiz}}, \bibinfo {author} {\bibfnamefont {C.~L.}\ \bibnamefont {Melamed}},
  \bibinfo {author} {\bibfnamefont {L.~T.}\ \bibnamefont {Schelhas}}, \bibinfo
  {author} {\bibfnamefont {K.~H.}\ \bibnamefont {Stone}}, \bibinfo {author}
  {\bibfnamefont {M.~F.}\ \bibnamefont {Toney}}, \bibinfo {author}
  {\bibfnamefont {E.~S.}\ \bibnamefont {Toberer}},\ and\ \bibinfo {author}
  {\bibfnamefont {A.~C.}\ \bibnamefont {Tamboli}},\ }\bibfield  {title}
  {\bibinfo {title} {Using resonant energy x-ray diffraction to extract
  chemical order parameters in ternary semiconductors},\ }\href
  {https://doi.org/10.1039/C9TC06699C} {\bibfield  {journal} {\bibinfo
  {journal} {J. Mater. Chem. C}\ }\textbf {\bibinfo {volume} {8}},\ \bibinfo
  {pages} {4350} (\bibinfo {year} {2020}{\natexlab{b}})}\BibitemShut {NoStop}%
\bibitem [{\citenamefont {Christensen}\ \emph {et~al.}(2006)\citenamefont
  {Christensen}, \citenamefont {Lock}, \citenamefont {Overgaard},\ and\
  \citenamefont {Iversen}}]{doi:10.1021/ja063695y}%
  \BibitemOpen
  \bibfield  {author} {\bibinfo {author} {\bibfnamefont {M.}~\bibnamefont
  {Christensen}}, \bibinfo {author} {\bibfnamefont {N.}~\bibnamefont {Lock}},
  \bibinfo {author} {\bibfnamefont {J.}~\bibnamefont {Overgaard}},\ and\
  \bibinfo {author} {\bibfnamefont {B.~B.}\ \bibnamefont {Iversen}},\
  }\bibfield  {title} {\bibinfo {title} {Crystal structures of thermoelectric
  n- and p-type $\mathrm{Ba_8Ga_{16}Ge_{30}}$ studied by single crystal,
  multitemperature, neutron diffraction, conventional x-ray diffraction and
  resonant synchrotron x-ray diffraction},\ }\href
  {https://doi.org/10.1021/ja063695y} {\bibfield  {journal} {\bibinfo
  {journal} {Journal of the American Chemical Society}\ }\textbf {\bibinfo
  {volume} {128}},\ \bibinfo {pages} {15657} (\bibinfo {year} {2006})},\
  \bibinfo {note} {pMID: 17147375}\BibitemShut {NoStop}%
\bibitem [{\citenamefont {Stone}\ \emph {et~al.}(2016)\citenamefont {Stone},
  \citenamefont {Christensen}, \citenamefont {Harvey}, \citenamefont {Teeter},
  \citenamefont {Repins},\ and\ \citenamefont {Toney}}]{Kevin}%
  \BibitemOpen
  \bibfield  {author} {\bibinfo {author} {\bibfnamefont {K.~H.}\ \bibnamefont
  {Stone}}, \bibinfo {author} {\bibfnamefont {S.~T.}\ \bibnamefont
  {Christensen}}, \bibinfo {author} {\bibfnamefont {S.~P.}\ \bibnamefont
  {Harvey}}, \bibinfo {author} {\bibfnamefont {G.}~\bibnamefont {Teeter}},
  \bibinfo {author} {\bibfnamefont {I.~L.}\ \bibnamefont {Repins}},\ and\
  \bibinfo {author} {\bibfnamefont {M.~F.}\ \bibnamefont {Toney}},\ }\bibfield
  {title} {\bibinfo {title} {Quantifying point defects in
  $\mathrm{Cu_2ZnSn(S,Se)_4}$ thin films using resonant x-ray diffraction},\
  }\href {https://doi.org/10.1063/1.4964738} {\bibfield  {journal} {\bibinfo
  {journal} {Applied Physics Letters}\ }\textbf {\bibinfo {volume} {109}},\
  \bibinfo {pages} {161901} (\bibinfo {year} {2016})}\BibitemShut {NoStop}%
\bibitem [{\citenamefont {Schelhas}\ \emph {et~al.}(2017)\citenamefont
  {Schelhas}, \citenamefont {Stone}, \citenamefont {Harvey}, \citenamefont
  {Zakhidov}, \citenamefont {Salleo}, \citenamefont {Teeter}, \citenamefont
  {Repins},\ and\ \citenamefont {Toney}}]{Laura}%
  \BibitemOpen
  \bibfield  {author} {\bibinfo {author} {\bibfnamefont {L.~T.}\ \bibnamefont
  {Schelhas}}, \bibinfo {author} {\bibfnamefont {K.~H.}\ \bibnamefont {Stone}},
  \bibinfo {author} {\bibfnamefont {S.~P.}\ \bibnamefont {Harvey}}, \bibinfo
  {author} {\bibfnamefont {D.}~\bibnamefont {Zakhidov}}, \bibinfo {author}
  {\bibfnamefont {A.}~\bibnamefont {Salleo}}, \bibinfo {author} {\bibfnamefont
  {G.}~\bibnamefont {Teeter}}, \bibinfo {author} {\bibfnamefont {I.~L.}\
  \bibnamefont {Repins}},\ and\ \bibinfo {author} {\bibfnamefont {M.~F.}\
  \bibnamefont {Toney}},\ }\bibfield  {title} {\bibinfo {title} {Point defects
  in $\mathrm{Cu_2ZnSnSe_4 (CZTSe)}$: Resonant x-ray diffraction study of the
  low-temperature order/disorder transition},\ }\href
  {https://doi.org/10.1002/pssb.201700156} {\bibfield  {journal} {\bibinfo
  {journal} {Physica Status Solidi (b)}\ }\textbf {\bibinfo {volume} {254}},\
  \bibinfo {pages} {1700156} (\bibinfo {year} {2017})}\BibitemShut {NoStop}%
\bibitem [{\citenamefont {Dmitrienko}\ and\ \citenamefont
  {Ovchinnikova}(2000)}]{Dmitrienko:av0029}%
  \BibitemOpen
  \bibfield  {author} {\bibinfo {author} {\bibfnamefont {V.~E.}\ \bibnamefont
  {Dmitrienko}}\ and\ \bibinfo {author} {\bibfnamefont {E.~N.}\ \bibnamefont
  {Ovchinnikova}},\ }\bibfield  {title} {\bibinfo {title} {{Resonant X-ray
  diffraction: `forbidden' Bragg reflections induced by thermal vibrations and
  point defects}},\ }\href {https://doi.org/10.1107/S0108767300003421}
  {\bibfield  {journal} {\bibinfo  {journal} {Acta Crystallographica Section
  A}\ }\textbf {\bibinfo {volume} {56}},\ \bibinfo {pages} {340} (\bibinfo
  {year} {2000})}\BibitemShut {NoStop}%
\bibitem [{\citenamefont {Ortiz}\ \emph {et~al.}(2018)\citenamefont {Ortiz},
  \citenamefont {Peng}, \citenamefont {Gomes}, \citenamefont {Gorai},
  \citenamefont {Zhu}, \citenamefont {Smiadak}, \citenamefont {Snyder},
  \citenamefont {Stevanovic}, \citenamefont {Ertekin}, \citenamefont
  {Zevalkink},\ and\ \citenamefont {Toberer}}]{Brenden1}%
  \BibitemOpen
  \bibfield  {author} {\bibinfo {author} {\bibfnamefont {B.~R.}\ \bibnamefont
  {Ortiz}}, \bibinfo {author} {\bibfnamefont {W.}~\bibnamefont {Peng}},
  \bibinfo {author} {\bibfnamefont {L.~C.}\ \bibnamefont {Gomes}}, \bibinfo
  {author} {\bibfnamefont {P.}~\bibnamefont {Gorai}}, \bibinfo {author}
  {\bibfnamefont {T.}~\bibnamefont {Zhu}}, \bibinfo {author} {\bibfnamefont
  {D.~M.}\ \bibnamefont {Smiadak}}, \bibinfo {author} {\bibfnamefont {G.~J.}\
  \bibnamefont {Snyder}}, \bibinfo {author} {\bibfnamefont {V.}~\bibnamefont
  {Stevanovic}}, \bibinfo {author} {\bibfnamefont {E.}~\bibnamefont {Ertekin}},
  \bibinfo {author} {\bibfnamefont {A.}~\bibnamefont {Zevalkink}},\ and\
  \bibinfo {author} {\bibfnamefont {E.~S.}\ \bibnamefont {Toberer}},\
  }\bibfield  {title} {\bibinfo {title} {Ultralow thermal conductivity in
  diamond-like semiconductors: Selective scattering of phonons from antisite
  defects},\ }\href {https://doi.org/10.1021/acs.chemmater.8b00890} {\bibfield
  {journal} {\bibinfo  {journal} {Chemistry of Materials}\ }\textbf {\bibinfo
  {volume} {30}},\ \bibinfo {pages} {3395} (\bibinfo {year}
  {2018})}\BibitemShut {NoStop}%
\bibitem [{\citenamefont {Hsieh}\ \emph {et~al.}(2016)\citenamefont {Hsieh},
  \citenamefont {Han}, \citenamefont {Jiang}, \citenamefont {Song},
  \citenamefont {Chen}, \citenamefont {Meng}, \citenamefont {Zhou},\ and\
  \citenamefont {Yang}}]{doi:10.1002/aenm.201502386}%
  \BibitemOpen
  \bibfield  {author} {\bibinfo {author} {\bibfnamefont {Y.-T.}\ \bibnamefont
  {Hsieh}}, \bibinfo {author} {\bibfnamefont {Q.}~\bibnamefont {Han}}, \bibinfo
  {author} {\bibfnamefont {C.}~\bibnamefont {Jiang}}, \bibinfo {author}
  {\bibfnamefont {T.-B.}\ \bibnamefont {Song}}, \bibinfo {author}
  {\bibfnamefont {H.}~\bibnamefont {Chen}}, \bibinfo {author} {\bibfnamefont
  {L.}~\bibnamefont {Meng}}, \bibinfo {author} {\bibfnamefont {H.}~\bibnamefont
  {Zhou}},\ and\ \bibinfo {author} {\bibfnamefont {Y.}~\bibnamefont {Yang}},\
  }\bibfield  {title} {\bibinfo {title} {Efficiency enhancement of
  $\mathrm{Cu_2ZnSn(S,Se)_4}$ solar cells via alkali metals doping},\ }\href
  {https://doi.org/10.1002/aenm.201502386} {\bibfield  {journal} {\bibinfo
  {journal} {Advanced Energy Materials}\ }\textbf {\bibinfo {volume} {6}},\
  \bibinfo {pages} {1502386} (\bibinfo {year} {2016})}\BibitemShut {NoStop}%
\bibitem [{\citenamefont {Cheng}\ \emph {et~al.}(2014)\citenamefont {Cheng},
  \citenamefont {Liu}, \citenamefont {Bai}, \citenamefont {Shi},\ and\
  \citenamefont {Chen}}]{doi:10.1063/1.4872250}%
  \BibitemOpen
  \bibfield  {author} {\bibinfo {author} {\bibfnamefont {N.}~\bibnamefont
  {Cheng}}, \bibinfo {author} {\bibfnamefont {R.}~\bibnamefont {Liu}}, \bibinfo
  {author} {\bibfnamefont {S.}~\bibnamefont {Bai}}, \bibinfo {author}
  {\bibfnamefont {X.}~\bibnamefont {Shi}},\ and\ \bibinfo {author}
  {\bibfnamefont {L.}~\bibnamefont {Chen}},\ }\bibfield  {title} {\bibinfo
  {title} {Enhanced thermoelectric performance in $\mathrm{Cd}$ doped
  $\mathrm{CuInTe_2}$ compounds},\ }\href {https://doi.org/10.1063/1.4872250}
  {\bibfield  {journal} {\bibinfo  {journal} {Journal of Applied Physics}\
  }\textbf {\bibinfo {volume} {115}},\ \bibinfo {pages} {163705} (\bibinfo
  {year} {2014})}\BibitemShut {NoStop}%
\bibitem [{\citenamefont {Kosuga}\ \emph
  {et~al.}(2012{\natexlab{a}})\citenamefont {Kosuga}, \citenamefont
  {Higashine}, \citenamefont {Plirdpring}, \citenamefont {Matsuzawa},
  \citenamefont {Kurosaki},\ and\ \citenamefont {Yamanaka}}]{Kosuga_2012}%
  \BibitemOpen
  \bibfield  {author} {\bibinfo {author} {\bibfnamefont {A.}~\bibnamefont
  {Kosuga}}, \bibinfo {author} {\bibfnamefont {R.}~\bibnamefont {Higashine}},
  \bibinfo {author} {\bibfnamefont {T.}~\bibnamefont {Plirdpring}}, \bibinfo
  {author} {\bibfnamefont {M.}~\bibnamefont {Matsuzawa}}, \bibinfo {author}
  {\bibfnamefont {K.}~\bibnamefont {Kurosaki}},\ and\ \bibinfo {author}
  {\bibfnamefont {S.}~\bibnamefont {Yamanaka}},\ }\bibfield  {title} {\bibinfo
  {title} {Effects of the defects on the thermoelectric properties of
  $\mathrm{Cu{\textendash}In{\textendash}Te}$ chalcopyrite-related compounds},\
  }\href {https://doi.org/10.1143/jjap.51.121803} {\bibfield  {journal}
  {\bibinfo  {journal} {Japanese Journal of Applied Physics}\ }\textbf
  {\bibinfo {volume} {51}},\ \bibinfo {pages} {121803} (\bibinfo {year}
  {2012}{\natexlab{a}})}\BibitemShut {NoStop}%
\bibitem [{\citenamefont {Kosuga}\ \emph
  {et~al.}(2012{\natexlab{b}})\citenamefont {Kosuga}, \citenamefont
  {Plirdpring}, \citenamefont {Higashine}, \citenamefont {Matsuzawa},
  \citenamefont {Kurosaki},\ and\ \citenamefont
  {Yamanaka}}]{doi:10.1063/1.3678044}%
  \BibitemOpen
  \bibfield  {author} {\bibinfo {author} {\bibfnamefont {A.}~\bibnamefont
  {Kosuga}}, \bibinfo {author} {\bibfnamefont {T.}~\bibnamefont {Plirdpring}},
  \bibinfo {author} {\bibfnamefont {R.}~\bibnamefont {Higashine}}, \bibinfo
  {author} {\bibfnamefont {M.}~\bibnamefont {Matsuzawa}}, \bibinfo {author}
  {\bibfnamefont {K.}~\bibnamefont {Kurosaki}},\ and\ \bibinfo {author}
  {\bibfnamefont {S.}~\bibnamefont {Yamanaka}},\ }\bibfield  {title} {\bibinfo
  {title} {High-temperature thermoelectric properties of
  $\mathrm{Cu_{1–x}InTe_2}$ with a chalcopyrite structure},\ }\href
  {https://doi.org/10.1063/1.3678044} {\bibfield  {journal} {\bibinfo
  {journal} {Applied Physics Letters}\ }\textbf {\bibinfo {volume} {100}},\
  \bibinfo {pages} {042108} (\bibinfo {year} {2012}{\natexlab{b}})}\BibitemShut
  {NoStop}%
\bibitem [{\citenamefont {Yusufu}\ \emph {et~al.}(2011)\citenamefont {Yusufu},
  \citenamefont {Kurosaki}, \citenamefont {Kosuga}, \citenamefont {Sugahara},
  \citenamefont {Ohishi}, \citenamefont {Muta},\ and\ \citenamefont
  {Yamanaka}}]{doi:10.1063/1.3617458}%
  \BibitemOpen
  \bibfield  {author} {\bibinfo {author} {\bibfnamefont {A.}~\bibnamefont
  {Yusufu}}, \bibinfo {author} {\bibfnamefont {K.}~\bibnamefont {Kurosaki}},
  \bibinfo {author} {\bibfnamefont {A.}~\bibnamefont {Kosuga}}, \bibinfo
  {author} {\bibfnamefont {T.}~\bibnamefont {Sugahara}}, \bibinfo {author}
  {\bibfnamefont {Y.}~\bibnamefont {Ohishi}}, \bibinfo {author} {\bibfnamefont
  {H.}~\bibnamefont {Muta}},\ and\ \bibinfo {author} {\bibfnamefont
  {S.}~\bibnamefont {Yamanaka}},\ }\bibfield  {title} {\bibinfo {title}
  {Thermoelectric properties of $\mathrm{Ag_{1−x}GaTe_2}$ with chalcopyrite
  structure},\ }\href {https://doi.org/10.1063/1.3617458} {\bibfield  {journal}
  {\bibinfo  {journal} {Applied Physics Letters}\ }\textbf {\bibinfo {volume}
  {99}},\ \bibinfo {pages} {061902} (\bibinfo {year} {2011})}\BibitemShut
  {NoStop}%
\bibitem [{\citenamefont {Xie}\ \emph {et~al.}(2017)\citenamefont {Xie},
  \citenamefont {Su}, \citenamefont {Zheng}, \citenamefont {Zhu}, \citenamefont
  {Yin}, \citenamefont {Yan}, \citenamefont {Uher}, \citenamefont
  {Kanatzidis},\ and\ \citenamefont {Tang}}]{doi:10.1002/aenm.201601299}%
  \BibitemOpen
  \bibfield  {author} {\bibinfo {author} {\bibfnamefont {H.}~\bibnamefont
  {Xie}}, \bibinfo {author} {\bibfnamefont {X.}~\bibnamefont {Su}}, \bibinfo
  {author} {\bibfnamefont {G.}~\bibnamefont {Zheng}}, \bibinfo {author}
  {\bibfnamefont {T.}~\bibnamefont {Zhu}}, \bibinfo {author} {\bibfnamefont
  {K.}~\bibnamefont {Yin}}, \bibinfo {author} {\bibfnamefont {Y.}~\bibnamefont
  {Yan}}, \bibinfo {author} {\bibfnamefont {C.}~\bibnamefont {Uher}}, \bibinfo
  {author} {\bibfnamefont {M.~G.}\ \bibnamefont {Kanatzidis}},\ and\ \bibinfo
  {author} {\bibfnamefont {X.}~\bibnamefont {Tang}},\ }\bibfield  {title}
  {\bibinfo {title} {The role of $\mathrm{Zn}$ in chalcopyrite
  $\mathrm{CuFeS_2}$: Enhanced thermoelectric properties of
  $\mathrm{Cu_{1–x}Zn_xFeS_2}$ with in-situ nanoprecipitates},\ }\href
  {https://doi.org/10.1002/aenm.201601299} {\bibfield  {journal} {\bibinfo
  {journal} {Advanced Energy Materials}\ }\textbf {\bibinfo {volume} {7}},\
  \bibinfo {pages} {1601299} (\bibinfo {year} {2017})}\BibitemShut {NoStop}%
\bibitem [{\citenamefont {Li}\ \emph {et~al.}(2014)\citenamefont {Li},
  \citenamefont {Zhang}, \citenamefont {Qin}, \citenamefont {Day},
  \citenamefont {Jeffrey~Snyder}, \citenamefont {Shi},\ and\ \citenamefont
  {Chen}}]{doi:10.1063/1.4902849}%
  \BibitemOpen
  \bibfield  {author} {\bibinfo {author} {\bibfnamefont {Y.}~\bibnamefont
  {Li}}, \bibinfo {author} {\bibfnamefont {T.}~\bibnamefont {Zhang}}, \bibinfo
  {author} {\bibfnamefont {Y.}~\bibnamefont {Qin}}, \bibinfo {author}
  {\bibfnamefont {T.}~\bibnamefont {Day}}, \bibinfo {author} {\bibfnamefont
  {G.}~\bibnamefont {Jeffrey~Snyder}}, \bibinfo {author} {\bibfnamefont
  {X.}~\bibnamefont {Shi}},\ and\ \bibinfo {author} {\bibfnamefont
  {L.}~\bibnamefont {Chen}},\ }\bibfield  {title} {\bibinfo {title}
  {Thermoelectric transport properties of diamond-like
  $\mathrm{Cu_{1−x}Fe_{1+x}S_2}$ tetrahedral compounds},\ }\href
  {https://doi.org/10.1063/1.4902849} {\bibfield  {journal} {\bibinfo
  {journal} {Journal of Applied Physics}\ }\textbf {\bibinfo {volume} {116}},\
  \bibinfo {pages} {203705} (\bibinfo {year} {2014})}\BibitemShut {NoStop}%
\bibitem [{\citenamefont {Zhang}\ \emph {et~al.}(2017)\citenamefont {Zhang},
  \citenamefont {Yang}, \citenamefont {Jiang}, \citenamefont {Zhou},
  \citenamefont {Li}, \citenamefont {Xin}, \citenamefont {Basit}, \citenamefont
  {Ren},\ and\ \citenamefont {He}}]{ZHANG2017156}%
  \BibitemOpen
  \bibfield  {author} {\bibinfo {author} {\bibfnamefont {D.}~\bibnamefont
  {Zhang}}, \bibinfo {author} {\bibfnamefont {J.}~\bibnamefont {Yang}},
  \bibinfo {author} {\bibfnamefont {Q.}~\bibnamefont {Jiang}}, \bibinfo
  {author} {\bibfnamefont {Z.}~\bibnamefont {Zhou}}, \bibinfo {author}
  {\bibfnamefont {X.}~\bibnamefont {Li}}, \bibinfo {author} {\bibfnamefont
  {J.}~\bibnamefont {Xin}}, \bibinfo {author} {\bibfnamefont {A.}~\bibnamefont
  {Basit}}, \bibinfo {author} {\bibfnamefont {Y.}~\bibnamefont {Ren}},\ and\
  \bibinfo {author} {\bibfnamefont {X.}~\bibnamefont {He}},\ }\bibfield
  {title} {\bibinfo {title} {Multi-cations compound $\mathrm{Cu_2CoSnS_4}$: Dft
  calculating, band engineering and thermoelectric performance regulation},\
  }\href {https://doi.org/https://doi.org/10.1016/j.nanoen.2017.04.027}
  {\bibfield  {journal} {\bibinfo  {journal} {Nano Energy}\ }\textbf {\bibinfo
  {volume} {36}},\ \bibinfo {pages} {156 } (\bibinfo {year}
  {2017})}\BibitemShut {NoStop}%
\bibitem [{\citenamefont {Zhou}\ \emph {et~al.}(2017)\citenamefont {Zhou},
  \citenamefont {Chen}, \citenamefont {Dong},\ and\ \citenamefont
  {Yin}}]{Zhou:2017:1941-4900:1520}%
  \BibitemOpen
  \bibfield  {author} {\bibinfo {author} {\bibfnamefont {Y.}~\bibnamefont
  {Zhou}}, \bibinfo {author} {\bibfnamefont {Q.}~\bibnamefont {Chen}}, \bibinfo
  {author} {\bibfnamefont {L.}~\bibnamefont {Dong}},\ and\ \bibinfo {author}
  {\bibfnamefont {Y.}~\bibnamefont {Yin}},\ }\bibfield  {title} {\bibinfo
  {title} {Improving thermoelectric performance of chalcogenide
  $\mathrm{Cu_{2-2x}CdSnSe_4}$ by $\mathrm{Cu}$ vacancy},\ }\href
  {https://doi.org/doi:10.1166/nnl.2017.2518} {\bibfield  {journal} {\bibinfo
  {journal} {Nanoscience and Nanotechnology Letters}\ }\textbf {\bibinfo
  {volume} {9}},\ \bibinfo {pages} {1520} (\bibinfo {year} {2017})}\BibitemShut
  {NoStop}%
\bibitem [{\citenamefont {Chetty}\ \emph {et~al.}(2016)\citenamefont {Chetty},
  \citenamefont {Bali},\ and\ \citenamefont {Mallik}}]{CHETTY201617}%
  \BibitemOpen
  \bibfield  {author} {\bibinfo {author} {\bibfnamefont {R.}~\bibnamefont
  {Chetty}}, \bibinfo {author} {\bibfnamefont {A.}~\bibnamefont {Bali}},\ and\
  \bibinfo {author} {\bibfnamefont {R.~C.}\ \bibnamefont {Mallik}},\ }\bibfield
   {title} {\bibinfo {title} {Thermoelectric properties of indium doped
  $\mathrm{Cu_2CdSnSe_4}$},\ }\href
  {https://doi.org/https://doi.org/10.1016/j.intermet.2016.01.004} {\bibfield
  {journal} {\bibinfo  {journal} {Intermetallics}\ }\textbf {\bibinfo {volume}
  {72}},\ \bibinfo {pages} {17 } (\bibinfo {year} {2016})}\BibitemShut
  {NoStop}%
\bibitem [{\citenamefont {Liu}\ \emph {et~al.}(2009)\citenamefont {Liu},
  \citenamefont {Chen}, \citenamefont {Huang},\ and\ \citenamefont
  {Chen}}]{doi:10.1002/adma.200900409}%
  \BibitemOpen
  \bibfield  {author} {\bibinfo {author} {\bibfnamefont {M.-L.}\ \bibnamefont
  {Liu}}, \bibinfo {author} {\bibfnamefont {I.-W.}\ \bibnamefont {Chen}},
  \bibinfo {author} {\bibfnamefont {F.-Q.}\ \bibnamefont {Huang}},\ and\
  \bibinfo {author} {\bibfnamefont {L.-D.}\ \bibnamefont {Chen}},\ }\bibfield
  {title} {\bibinfo {title} {Improved thermoelectric properties of
  $\mathrm{Cu}$-doped quaternary chalcogenides of $\mathrm{Cu_2CdSnSe_4}$},\
  }\href {https://doi.org/10.1002/adma.200900409} {\bibfield  {journal}
  {\bibinfo  {journal} {Advanced Materials}\ }\textbf {\bibinfo {volume}
  {21}},\ \bibinfo {pages} {3808} (\bibinfo {year} {2009})}\BibitemShut
  {NoStop}%
\bibitem [{\citenamefont {Shi}\ \emph {et~al.}(2009)\citenamefont {Shi},
  \citenamefont {Huang}, \citenamefont {Liu},\ and\ \citenamefont
  {Chen}}]{doi:10.1063/1.3103604}%
  \BibitemOpen
  \bibfield  {author} {\bibinfo {author} {\bibfnamefont {X.~Y.}\ \bibnamefont
  {Shi}}, \bibinfo {author} {\bibfnamefont {F.~Q.}\ \bibnamefont {Huang}},
  \bibinfo {author} {\bibfnamefont {M.~L.}\ \bibnamefont {Liu}},\ and\ \bibinfo
  {author} {\bibfnamefont {L.~D.}\ \bibnamefont {Chen}},\ }\bibfield  {title}
  {\bibinfo {title} {Thermoelectric properties of tetrahedrally bonded wide-gap
  stannite compounds $\mathrm{Cu_2ZnSn_{1−x}In_xSe_4}$},\ }\href
  {https://doi.org/10.1063/1.3103604} {\bibfield  {journal} {\bibinfo
  {journal} {Applied Physics Letters}\ }\textbf {\bibinfo {volume} {94}},\
  \bibinfo {pages} {122103} (\bibinfo {year} {2009})}\BibitemShut {NoStop}%
\bibitem [{\citenamefont {Ortiz}\ \emph {et~al.}(2019)\citenamefont {Ortiz},
  \citenamefont {Gordiz}, \citenamefont {Gomes}, \citenamefont {Braden},
  \citenamefont {Adamczyk}, \citenamefont {Qu}, \citenamefont {Ertekin},\ and\
  \citenamefont {Toberer}}]{C8TA10332A}%
  \BibitemOpen
  \bibfield  {author} {\bibinfo {author} {\bibfnamefont {B.~R.}\ \bibnamefont
  {Ortiz}}, \bibinfo {author} {\bibfnamefont {K.}~\bibnamefont {Gordiz}},
  \bibinfo {author} {\bibfnamefont {L.~C.}\ \bibnamefont {Gomes}}, \bibinfo
  {author} {\bibfnamefont {T.}~\bibnamefont {Braden}}, \bibinfo {author}
  {\bibfnamefont {J.~M.}\ \bibnamefont {Adamczyk}}, \bibinfo {author}
  {\bibfnamefont {J.}~\bibnamefont {Qu}}, \bibinfo {author} {\bibfnamefont
  {E.}~\bibnamefont {Ertekin}},\ and\ \bibinfo {author} {\bibfnamefont {E.~S.}\
  \bibnamefont {Toberer}},\ }\bibfield  {title} {\bibinfo {title} {Carrier
  density control in $\mathrm{Cu_2HgGeTe_4}$ and discovery of
  $\mathrm{Hg_2GeTe_4}$ via phase boundary mapping},\ }\href
  {https://doi.org/10.1039/C8TA10332A} {\bibfield  {journal} {\bibinfo
  {journal} {J. Mater. Chem. A}\ }\textbf {\bibinfo {volume} {7}},\ \bibinfo
  {pages} {621} (\bibinfo {year} {2019})}\BibitemShut {NoStop}%
\bibitem [{\citenamefont {Efron}\ and\ \citenamefont
  {Tibshirani}(1986)}]{efron1986}%
  \BibitemOpen
  \bibfield  {author} {\bibinfo {author} {\bibfnamefont {B.}~\bibnamefont
  {Efron}}\ and\ \bibinfo {author} {\bibfnamefont {R.}~\bibnamefont
  {Tibshirani}},\ }\bibfield  {title} {\bibinfo {title} {Bootstrap methods for
  standard errors, confidence intervals, and other measures of statistical
  accuracy},\ }\href {https://doi.org/10.1214/ss/1177013815} {\bibfield
  {journal} {\bibinfo  {journal} {Statist. Sci.}\ }\textbf {\bibinfo {volume}
  {1}},\ \bibinfo {pages} {54} (\bibinfo {year} {1986})}\BibitemShut {NoStop}%
\bibitem [{\citenamefont {DiCiccio}\ and\ \citenamefont
  {Efron}(1996)}]{diciccio1996}%
  \BibitemOpen
  \bibfield  {author} {\bibinfo {author} {\bibfnamefont {T.~J.}\ \bibnamefont
  {DiCiccio}}\ and\ \bibinfo {author} {\bibfnamefont {B.}~\bibnamefont
  {Efron}},\ }\bibfield  {title} {\bibinfo {title} {Bootstrap confidence
  intervals},\ }\href {https://doi.org/10.1214/ss/1032280214} {\bibfield
  {journal} {\bibinfo  {journal} {Statist. Sci.}\ }\textbf {\bibinfo {volume}
  {11}},\ \bibinfo {pages} {189} (\bibinfo {year} {1996})}\BibitemShut
  {NoStop}%
\bibitem [{\citenamefont {Chernick}(2007)}]{chernick_2007}%
  \BibitemOpen
  \bibfield  {author} {\bibinfo {author} {\bibfnamefont {M.~R.}\ \bibnamefont
  {Chernick}},\ }\href {https://doi.org/10.1002/9780470192573} {\emph {\bibinfo
  {title} {Bootstrap Methods: A Guide for Practitioners and Researchers}}},\
  \bibinfo {edition} {2nd}\ ed.\ (\bibinfo  {publisher} {John Wiley \& Sons,
  Ltd},\ \bibinfo {year} {2007})\BibitemShut {NoStop}%
\bibitem [{\citenamefont {Stephens}(1999)}]{Stephens:hn0085}%
  \BibitemOpen
  \bibfield  {author} {\bibinfo {author} {\bibfnamefont {P.~W.}\ \bibnamefont
  {Stephens}},\ }\bibfield  {title} {\bibinfo {title} {{Phenomenological model
  of anisotropic peak broadening in powder diffraction}},\ }\href
  {https://doi.org/10.1107/S0021889898006001} {\bibfield  {journal} {\bibinfo
  {journal} {Journal of Applied Crystallography}\ }\textbf {\bibinfo {volume}
  {32}},\ \bibinfo {pages} {281} (\bibinfo {year} {1999})}\BibitemShut
  {NoStop}%
\bibitem [{\citenamefont {Suortti}(1972)}]{Suortti:a09574}%
  \BibitemOpen
  \bibfield  {author} {\bibinfo {author} {\bibfnamefont {P.}~\bibnamefont
  {Suortti}},\ }\bibfield  {title} {\bibinfo {title} {{Effects of porosity and
  surface roughness on the X-ray intensity reflected from a powder specimen}},\
  }\href {https://doi.org/10.1107/S0021889872009707} {\bibfield  {journal}
  {\bibinfo  {journal} {Journal of Applied Crystallography}\ }\textbf {\bibinfo
  {volume} {5}},\ \bibinfo {pages} {325} (\bibinfo {year} {1972})}\BibitemShut
  {NoStop}%
\bibitem [{\citenamefont {Hohenberg}\ and\ \citenamefont
  {Kohn}(1964)}]{hohenberg-kohn-1964}%
  \BibitemOpen
  \bibfield  {author} {\bibinfo {author} {\bibfnamefont {P.}~\bibnamefont
  {Hohenberg}}\ and\ \bibinfo {author} {\bibfnamefont {W.}~\bibnamefont
  {Kohn}},\ }\bibfield  {title} {\bibinfo {title} {Inhomogeneous electron
  gas},\ }\href {https://doi.org/10.1103/PhysRev.136.B864} {\bibfield
  {journal} {\bibinfo  {journal} {Phys. Rev.}\ }\textbf {\bibinfo {volume}
  {136}},\ \bibinfo {pages} {B864} (\bibinfo {year} {1964})}\BibitemShut
  {NoStop}%
\bibitem [{\citenamefont {Kohn}\ and\ \citenamefont
  {Sham}(1965)}]{kohn-sham-1965}%
  \BibitemOpen
  \bibfield  {author} {\bibinfo {author} {\bibfnamefont {W.}~\bibnamefont
  {Kohn}}\ and\ \bibinfo {author} {\bibfnamefont {L.~J.}\ \bibnamefont
  {Sham}},\ }\bibfield  {title} {\bibinfo {title} {Self-consistent equations
  including exchange and correlation effects},\ }\href
  {https://doi.org/10.1103/PhysRev.140.A1133} {\bibfield  {journal} {\bibinfo
  {journal} {Phys. Rev.}\ }\textbf {\bibinfo {volume} {140}},\ \bibinfo {pages}
  {A1133} (\bibinfo {year} {1965})}\BibitemShut {NoStop}%
\bibitem [{\citenamefont {Kresse}\ and\ \citenamefont
  {Furthm\"uller}(1996)}]{kressePRB1996}%
  \BibitemOpen
  \bibfield  {author} {\bibinfo {author} {\bibfnamefont {G.}~\bibnamefont
  {Kresse}}\ and\ \bibinfo {author} {\bibfnamefont {J.}~\bibnamefont
  {Furthm\"uller}},\ }\bibfield  {title} {\bibinfo {title} {Efficient iterative
  schemes for ab initio total-energy calculations using a plane-wave basis
  set},\ }\href {https://doi.org/10.1103/PhysRevB.54.11169} {\bibfield
  {journal} {\bibinfo  {journal} {Phys. Rev. B}\ }\textbf {\bibinfo {volume}
  {54}},\ \bibinfo {pages} {11169} (\bibinfo {year} {1996})}\BibitemShut
  {NoStop}%
\bibitem [{\citenamefont {Perdew}\ \emph {et~al.}(1996)\citenamefont {Perdew},
  \citenamefont {Burke},\ and\ \citenamefont {Ernzerhof}}]{Perdew1996}%
  \BibitemOpen
  \bibfield  {author} {\bibinfo {author} {\bibfnamefont {J.~P.}\ \bibnamefont
  {Perdew}}, \bibinfo {author} {\bibfnamefont {K.}~\bibnamefont {Burke}},\ and\
  \bibinfo {author} {\bibfnamefont {M.}~\bibnamefont {Ernzerhof}},\ }\bibfield
  {title} {\bibinfo {title} {Generalized gradient approximation made simple},\
  }\href {https://doi.org/10.1103/PhysRevLett.77.3865} {\bibfield  {journal}
  {\bibinfo  {journal} {Phys. Rev. Lett.}\ }\textbf {\bibinfo {volume} {77}},\
  \bibinfo {pages} {3865} (\bibinfo {year} {1996})}\BibitemShut {NoStop}%
\bibitem [{\citenamefont {Bl\"ochl}(1994)}]{blochlPRB1994}%
  \BibitemOpen
  \bibfield  {author} {\bibinfo {author} {\bibfnamefont {P.~E.}\ \bibnamefont
  {Bl\"ochl}},\ }\bibfield  {title} {\bibinfo {title} {Projector augmented-wave
  method},\ }\href {https://doi.org/10.1103/PhysRevB.50.17953} {\bibfield
  {journal} {\bibinfo  {journal} {Phys. Rev. B}\ }\textbf {\bibinfo {volume}
  {50}},\ \bibinfo {pages} {17953} (\bibinfo {year} {1994})}\BibitemShut
  {NoStop}%
\bibitem [{\citenamefont {Monkhorst}\ and\ \citenamefont
  {Pack}(1976)}]{mp-kgrid}%
  \BibitemOpen
  \bibfield  {author} {\bibinfo {author} {\bibfnamefont {H.~J.}\ \bibnamefont
  {Monkhorst}}\ and\ \bibinfo {author} {\bibfnamefont {J.~D.}\ \bibnamefont
  {Pack}},\ }\bibfield  {title} {\bibinfo {title} {Special points for
  brillouin-zone integrations},\ }\href
  {https://doi.org/10.1103/PhysRevB.13.5188} {\bibfield  {journal} {\bibinfo
  {journal} {Phys. Rev. B}\ }\textbf {\bibinfo {volume} {13}},\ \bibinfo
  {pages} {5188} (\bibinfo {year} {1976})}\BibitemShut {NoStop}%
\bibitem [{CAS()}]{CASM}%
  \BibitemOpen
  \href@noop {} {}\bibinfo {note} {CASM, v0.2.1 (2017). Available from
  https://github.com/prisms-center/CASMcode. doi:
  10.5281/zenodo.546148}\BibitemShut {NoStop}%
\bibitem [{\citenamefont {Qu}\ \emph {et~al.}(2020)\citenamefont {Qu},
  \citenamefont {Gomes}, \citenamefont {Adamczyk}, \citenamefont {Toberer},\
  and\ \citenamefont {Ertekin}}]{Jiaxing}%
  \BibitemOpen
  \bibfield  {author} {\bibinfo {author} {\bibfnamefont {J.}~\bibnamefont
  {Qu}}, \bibinfo {author} {\bibfnamefont {L.~C.}\ \bibnamefont {Gomes}},
  \bibinfo {author} {\bibfnamefont {J.~M.}\ \bibnamefont {Adamczyk}}, \bibinfo
  {author} {\bibfnamefont {E.~S.}\ \bibnamefont {Toberer}},\ and\ \bibinfo
  {author} {\bibfnamefont {E.}~\bibnamefont {Ertekin}},\ }\bibfield  {title}
  {\bibinfo {title} {Dopability and carrier density control for thermoelectric
  applications}} (\bibinfo {year} {2020}),\ \bibinfo {note} {unpublished
  Manuscript}\BibitemShut {NoStop}%
\end{thebibliography}%

\end{document}